\documentclass[12pt]{article} 
\input epsf.tex 
\def\be{\begin{equation}} \def\ee{\end{equation}}
\def\bi{\begin{itemize}} \def\ei{\end{itemize}}
\def\bea{\begin{eqnarray}} \def\eea{\end{eqnarray}} \def\ba{\begin{array}}
\def\ea{\end{array}} \def\ben{\begin{enumerate}} \def\een{\end{enumerate}}
  
\def\lll{\label}
\newcommand{\eqn}[1]{(\ref{#1})}

\newcommand{\prl}[3]{Phys. Rev. Lett. {\bf#1} ({#2}) {#3}}

\newcommand{\hepth}[1]{{\tt arXiv:{#1}[hep-th]}}
\newcommand{\arxiv}[1]{{\tt arXiv:{#1}[hep-th]}}

\def\ep{\epsilon}

\def\br{\nonumber\\}

\begin{document}
{}~
\hfill \vbox{
\hbox{arXiv:2505.nnnn} 
\hbox{May'25}}
\break

\vskip 3.5cm
\centerline{\large \bf
Exact islands scenario for CFT systems}
 
\centerline{\large \bf
and}
 
\centerline{\large \bf
 critical ratios in higher geometry}
 
\vskip 1cm

\vspace*{1cm}

\centerline{\bf  Harvendra Singh }

\vspace*{.5cm}
\centerline{ \it  Theory Division, Saha Institute of Nuclear Physics} 
\centerline{ \it  1/AF Bidhannagar, Kolkata 700064, India}
\vspace*{.25cm}

\centerline{ \it  Homi Bhabha National Institute (HBNI)} 
\centerline{ \it  Anushaktinagar, Mumbai 400094, India}
\vspace*{.25cm}

\vspace*{.5cm}

\vskip.5cm


\centerline{\bf Abstract} \bigskip

We study $CFT_d$ systems which are
in contact with each other and  symmetrically arranged.
The system-B is treated as bath that surrounds system-A in the middle.
Our focus is to learn how the entanglement
entropy of a bath pair system changes as a function of its size. 
The total size of systems A and B taken together is kept fixed in this process. 
It is found that for strip shaped systems the bath entropy 
becomes maximum when respective system sizes follow Fibonacci type 
critical ratio condition. Beyond critical point when bath size
increases the bath entropy starts decreasing, where
 island and icebergs entropies
play important role. 
Interestingly entire effect of icebergs can be resummed 
giving rise to 'exact island' scenario for $CFT_d$ with $d>2$. 
Post criticality we also find important identity involving entropy differences
$S[B]-S[A]=S_l-S_{island}$ where island contribution is exact. 
The mutual information of far separated bath pair 
follows specific law $I(B:B) \propto {b^2\over (Distance)^d}$. 
It never vanishes for finite systems. 
Once system-A size approaches to Kaluza-Klein scale the bath 
entropy becomes discretized. In summary knowing island corrections
is vital for large bath entanglement entropy.

\vfill 
\eject

\baselineskip=16.2pt


\section{Introduction}

 The  AdS/CFT holographic duality  \cite{malda} has 
provided big insights to our understanding of entanglement in
strongly coupled quantum theories. 
We focus here  on various aspects of the entanglement between 
similar looking quantum subsystems those having common interfaces. 
Generally it is believed that sharing of 
quantum information between physical systems 
is guided by unitarity and locality.   
Under this quantum principle the basic understanding of the 
formation of black hole like states 
evolving from a loosely bound pure quantum matter phase 
and its subsequent
evaporation via Hawking radiation still remains an 
unsolved mystery. Though
it is believed that the evolution process would  be unitary and all
information inside the black hole interior which is hidden behind the horizon
would be recoverable once the black hole fully evaporates. Related to this 
aspect  there has been a proposal that the entanglement entropy curve for the
bath radiation should bend when the half Page-time is crossed \cite{page}. 
This is certainly true when a pure quantum system is divided 
into two smaller subsystems. 
But for mixed states involving hawking radiation, 
or for finite temperature CFTs dual of  AdS-black holes,
it is not that straight forward to obtain the Page curve under time evolution.  
However, some progress has been made recently in specific  models 
by coupling a near-CFT (NCFT) to an external radiation bath system (CFT), 
and in  other examples by involving  nonperturbative  techniques such as 
wormholes, replicas and islands \cite{almheri,replica19}. Some
answers to these difficult questions have been attempted.
\footnote{ Also find a review on information paradox along
different paradigms \cite{raju}
and other related references therein; 
see also [\cite{susski}-\cite{hashi}].} 
Especially for Karch-Randall braneworld
set up of gravity coupled to a bath QFT system one can also find a
massive island-ic formulation in
\cite{massiveislands}. Other interesting developments on entanglement
islands are explored in \cite{penington}

Particularly the  AMM proposal for generalized entanglement
entropy \cite{almheri} involves introduction of  entropy island $(I)$ 
contributions, and by including  gravitational entropy
 of respective island boundary 
$(\partial I)$. Accordingly in their proposal 
 the `quantum' entanglement entropy of  2-dimensional 
radiation bath subsystem $(B)$ can be expressed as
\bea\label{ficti1}
 S_{quantum}[B]=
\{ ext [ {Area(\partial I)\over 4G_N} + S[B ~U ~I]]\}_{min} 
\eea
This model uses a hybrid type  `gravity' plus 'gauge' theory
holographic set up. The model itself is primarily
based on QES proposal of \cite{engelhardt}. It includes contribution
of gravitational entropy of island's boundary 
to the bath entanglement entropy. 
The islands are usually disconnected
surfaces inside bulk 2d JT gravity \cite{JT,JT1}. 
The conformally AdS (or near AdS) JT-gravity is a holographic
dual theory of a point quantum system. 
The 1-dimensional QFT  lives on the boundary of a 
semi-infinite $CFT_2$ system (treated as a bath). 
In  low energy descriptions the JT gravity is treated as a system
which is in contact with  2-dim radiation bath living 
over flat Minkowskian
coordinate patches. So there are both field theoretic 
($S[B ~U ~I]$) as well as 
gravitational contributions (${Area(\partial I)\over 4G_N}$) 
present in the generalised entropy formulation eq.\eqn{ficti1}. 
Next there is a
 need to pick the lowest contribution 
out of a set of many such possible extrema, which 
 include  entropy of islands as well as radiation.
\footnote{In further extensions of the
hybrid models one also includes wormhole contributions, see \cite{replica19}.} 
Although complicated looking, the expression in \eqn{ficti1} 
seemingly reproduces a Page-curve for the radiation  entropy 
\cite{almheri}. However AMM proposal ignores contributions of other
subleading entropies which we altogether pronounced as '{\it icebergs}' 
entropies \cite{hs2022,hs2024}. We make it clear that
the icebergs are  contributions of various disconnected elements
to the bath entropy, very much like the islands area entropy. However
the iceberg entropies fall off
with the bath size  much faster as compared to the island entropy.
So these give only sub-sub leading contributions. In current work we
try to resum all  iceberg contributions when it is possible. 
This gives rise to {\bf Exact Island} scenario.

The important feature of AMM proposal is that it highlights 
the appearances of island entropy inside the bulk gravity in
 post Page transition phase.
The islands  are usually situated outside the horizon. 
The island could arise by means of some dynamical principle, 
 usually  associated with 
the presence of a large bath system that is in contact with the quantum dot. 
We focus on the aspect  that 
 subleading (disconnected) contributions of `iceberg' 
entropies would have to be properly accounted for in the bath system entropy.
Only in specific cases the icebergs can be resummed which gives rise to
{\bf exact island} 
The icebergs are present in 2-dimension CFT \cite{hs2022} as well as in 
 their higher dimensional $CFT_d$ cases \cite{hs2024}. We
emphasize that, in specific cases 
 no doubt there will exist island entropy but we must
also include (subleading) iceberg entropies as their net effect
 can be resummed! 
As pointed out the entropy of the B-A-B system, 
i.e. systems A and B together, as shown
in figure \eqn{fig23b},  
could be broken up, especially in large 
bath case \cite{hs2022,hs2024}, 
$$
 S[A \cup B]\equiv S[B]_{Pure} + S_{Island}+ S_{Icebergs}
$$
The right hand side of this equation contains perturbative expansion and is valid 
when system-B (as called surrounding bath system on either side) is sufficiently 
large enough compared to system-A (in the middle). This depends upon the criticality. 
Though icebergs contributions are highly suppressed due to their rapid fall off. 
 On the physical note the island and icebergs entropies arise mainly due to 
variety of quantum interactions between states of system-A and the system-B.
This  work is primarily motivated by an idea 
of resumming these contributions and 
determining exact islands when it is possible. 

\begin{figure}[h]
\centerline{\epsfxsize=2.5in
\epsffile{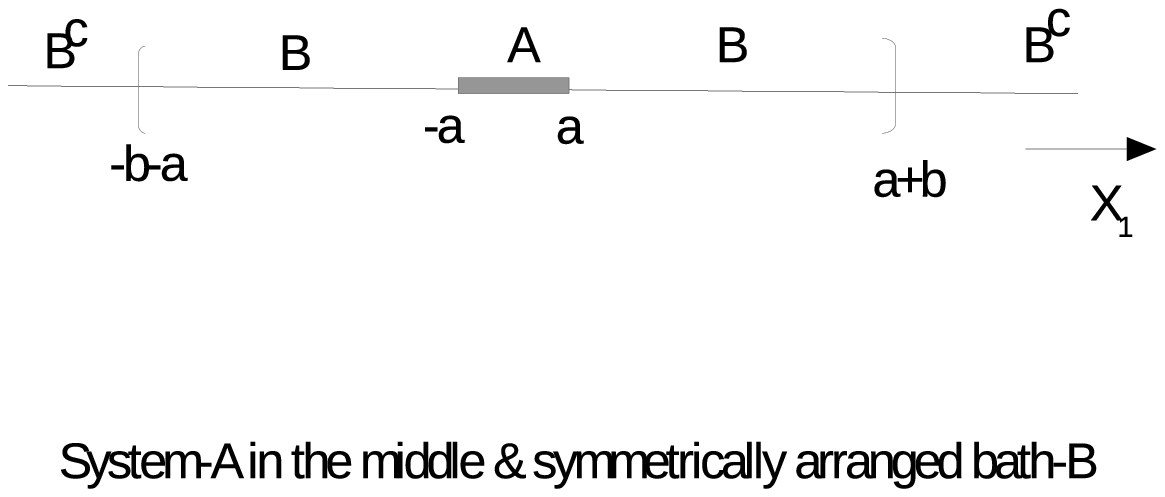} }
\caption{\label{fig23b} 
\small\it An arrangement of CFT system-A (along $x_1$ direction) 
and symmetrically arranged system-B (of size $b$ each situated on either side). 
The transverse spatial directions other than $x_1$ (if any) are all suppressed. 
The complimentary system $B^c=(A\cup B)^c$ is also indicated. One may treat
$B^c$ as the extended portion of the bath itself. }
\end{figure}

The rest of the article is organized as follows. In section-2 we explain 
how physical islands arise and the nature of their contributions 
to the bath entropy. We also discuss the mutual information of 
disjoint bath pair.
Section-3 contains how criticality of bath entropy arises in higher dimensional CFTs.
The ratios of critical sizes satisfy new identities which are  generalisation of
Golden ratio, involving natural
numbers as in Fibonacci-Pingala sequences. These also
involve higher dimensional geometries like squares and cubes etc.
The section-4 contains `exact' island proposal where all
icebergs contributions have been resummed. In 
post critical phase we find an identity involving entropy differences
$$S[B]-S[A]=S_l-S_{island}$$ 
where knowledge of exact island entropy plays crucial role, with out that the r.h.s. would be a constant.  
We also study the nature of local
part of bath entropy which is susceptible to small changes. 
The Kaluza-Klein discreteness of the 
entropy is highlighted in the section-5. The last section-6 has  summary
and discussions.

\section{Physical  islands}

Let us consider a system (A) in contact 
with a bath subsystem (B) in fully symmetrical set up 
as drawn in figure \eqn{fig23b}. 
Both systems have finite sizes and 
live on the boundary of the $AdS_{d+1}$ spacetime. 
Thus it is assumed that both systems A and B are made of 
identical field species, i.e. described by same field contents, 
for  simplicity of the problem. 
The pure $AdS_{d+1}$ spacetime geometry is described by
 following line element
\bea\label{ads3}
ds^2={L^2 \over z^2} (- dt^2 + dx_1^2 +\cdots + dx_{d-1}^2+ dz^2)
\eea
where constant $L$ represents a large radius 
of curvature (in string length units) 
of  AdS spacetime. 
The  coordinate ranges are taken as $-\infty\le (t,~x_i)\le\infty$ 
and  $0\le z \le \infty$, where coordinate $z$ 
represents the holographic range of boundary theory.
\begin{figure}[h]
\centerline{\epsfxsize=3in
\epsffile{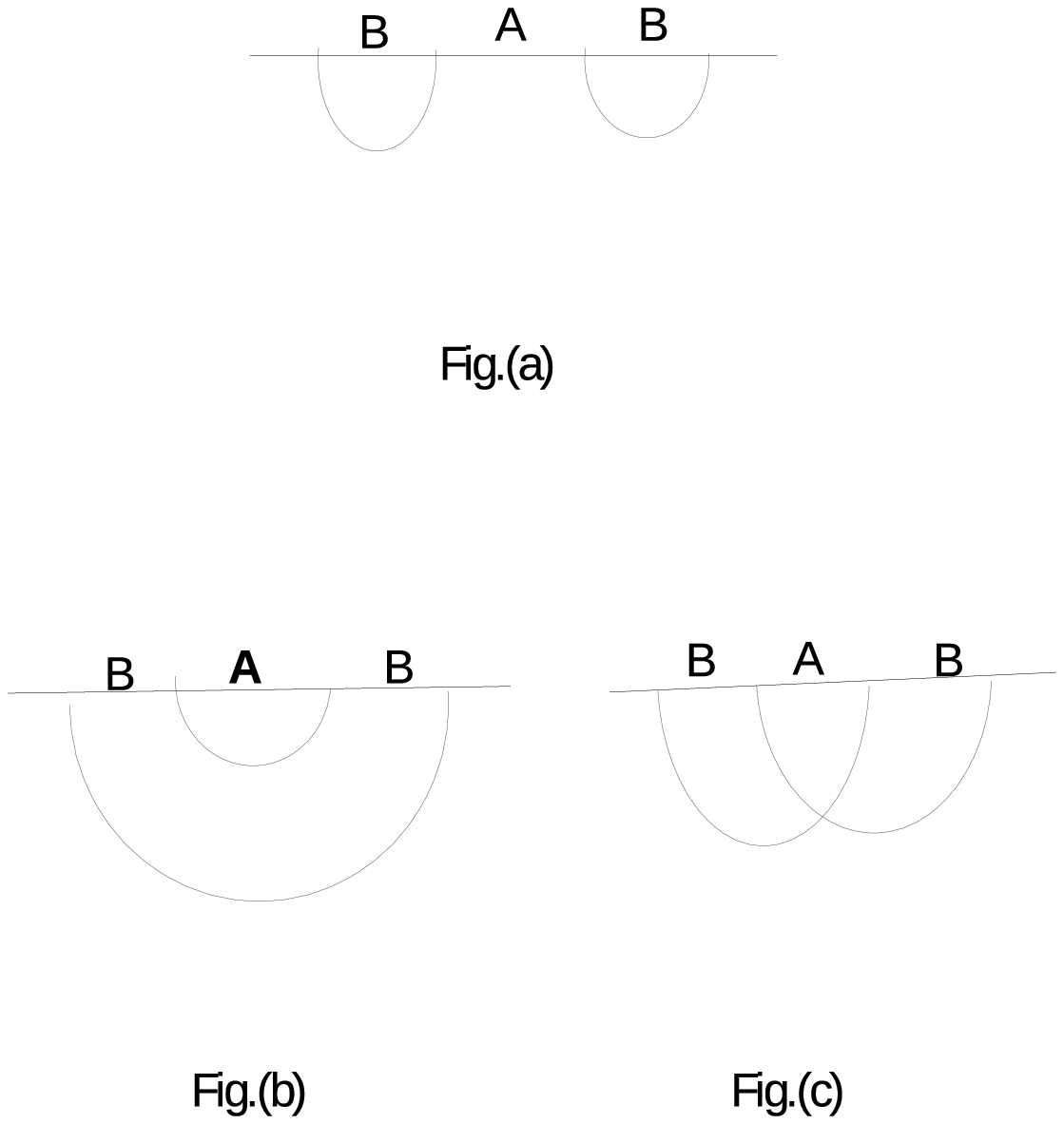} }
\caption{\label{fig22b} 
\small\it 
Three different types of RT surfaces are drawn schematically. 
The system-B (bath) is made of two disjoint intervals. 
All contribute to the system-B entropy.
When system-B is  small in size compared to system-A ($b\ll a$), the 
(disconnected) surfaces given in Fig.(a)  will give dominant contribution
to bath entropy. 
Two lower graphs instead have connecting components. 
When $b\gg a $, the Fig.(b) contributions dominate in the bath entropy.
The net system size $l$  same in all these graphs. 
}
\end{figure}

 \begin{figure}[h]
\centerline{\epsfxsize=3in
\epsffile{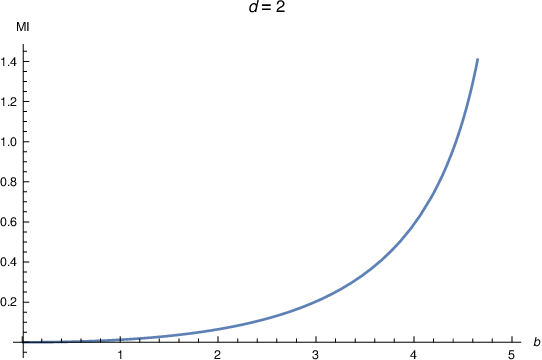} }
\caption{\label{fig24MI} 
\small\it 
The mutual information $I(B:B)$ between 
system-B pairs grows monotonically  with  size $b$, drawn for $CFT_2$ case. }
\end{figure} 

 \begin{figure}[h]
\centerline{\epsfxsize=3in
\epsffile{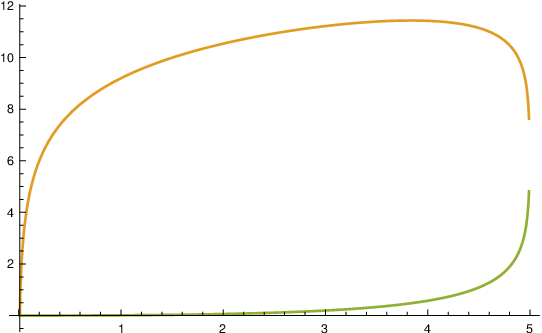} }
\caption{\label{fig24MIbh} 
\small\it 
In finite temperature case too the bath entropy $S[B]$
(upper curve) and the mutual information $I(B:B)$  (lower plot) smoothly
follow each other, for given fixed total size $l$. The entropy 
drops after getting to a maximum value at the critical point. 
Plot is shown here for thermal $CFT_2$. 
The horizontal axis represents half bath size $b$.
}
\end{figure} 

The $CFT_d$  lives on  $d$-dimensional  
$(t,\vec{x})$ flat Minkowski spacetime describing
boundary dynamics of $AdS_{d+1}$ bulk  geometry. 
Consider the set up in which a finite size bath system-B 
lives on the coordinate patches
$[-(b+a),-a]$ \& $[a, (b+a)]$, say
along $x_{1}$ direction, whereas 
system-A, sandwiched  between the bath, lives over the patch $[-a,a]$. 
The sketches are provided in figures \eqn{fig23b}, \eqn{fig22b} for clarity. 
The entire systems set up is arranged in  symmetrical way
for the convenience. The states of the system-A and the bath 
system-B are very much entangled. 
The complementary system $B^c=(A\cup B)^c$ lives over the coordinate patches 
$[b+a,\infty]$ and $[-(b+a),-\infty]$. 
The system $B^c$ is semi-infinite on either side.

From Ryu-Takayanagi holographic prescription \cite{RT,HRT}  the 
entanglement entropy of  a $CFT_d$ strip-shaped subsystem
 of width $l~(l=2a+2b)$ is given by \cite{hs2015}
\bea\label{tot1}
S(l)= 
{L^{d-1}V^{(d-2)} \over (2d-4) G_{d+1}}\left(
 {1\over \epsilon^{d-2}}-{b_0^{d-1}\over(a+b)^{d-2}}\right)
\equiv S_l
\eea
where  $\epsilon\simeq 0$ is the  UV 
cut-off  for  $CFT_d$ (we consider examples where $d>2$).
$V^{(d-2)}=l_2l_3\cdots l_{d-1}$ is the regulated large 
volume of the box along $(d-2)$ spatial directions
perpendicular to the strip width direction $x_{1}$.
\footnote{
The parameter 
$b_0={1\over 2(d-1)} B({d\over 2d-2},{1\over 2})$ is  specific 
$d$ dependent coefficient involving explicit 
Beta-functions;  more details in appendices of ref.\cite{hs2015}. 
The strip system lies in the range $-{l\over 2}\le x_1\le {l \over 2}$.} 
We will take $l$ sufficiently large but fixed, thus $ S(l)$  
has  constant value.
Our aim is to  determine entanglement entropy 
 when  $b$ is varied between $b= 0$ and $b= l/2$, by hand.  
We simply assume  conservation laws, so that the net gain (loss) 
of system-A is compensated by equal loss (gain) 
in the size of bath system-B 
and vice-versa. Note that such a mechanical process will keep $l$ fixed.  
({\it For explicit time dependent cases one
 may have some definite rate of change, or through an internally
 driven process,
while  $\dot a=-  \dot b$
would be true due to conservation of energy-momentum.
The exact rate of loss or gain and the mechanism by which it can happen
is not important to know here and the actual details of 
the physical process is also not required.}) 
All we are assuming is that
ordinary conservation laws are at work for  
complete system within fixed size $l$.\footnote{
Any explicit time dependent processes are not studied here.} 
Obviously we are assuming  that the
system-A and the bath-B are made up of 
identical fields (or quasi particles)
content. We consider two extreme cases below. 

We are interested in evaluating the HEE of a symmetrical bath system on either
side of the system-A. The complete bath system-B is made up 
of two disjoint intervals of same size $b$ each. 
The entanglement entropy of two identical intervals in $2d$ CFT
is  given by \cite{cftENT}
\bea\label{lko9}
S[B]&=&{L\over 2G_3} \log \left(2a l b^2\over\epsilon^2 (2a+b)^2\right) \br
\eea
which can also be expressed as
\bea\label{lko10}
S[B]&=&  S(b) +S(b)- I(B:B).
\eea
\footnote{ In the finite temperature cases the bath entropy becomes
$
S[B]={L\over 2G_3} \log \left(\sinh({2a\over z_0}) \sinh({l\over z_0}) 
\sinh^2({b\over z_0})\over
\sinh^2({2a+b\over z_0})\right)+S_{UV}$. The 
$z=z_0$ is the black hole horizon location in the $AdS$ space.  }
For our notation $S(b)$ is a function 
$S(b)={L\over 4G_3} \log({b^2\over \epsilon^2})$, it is the 
entropy of single interval of size $b$ in the $CFT_2$. Whereas  
mutual information $I(B:B)$ between two  bath patches is that 
part of the entropy
which effectively measures the connectivity between two disjoint patches. 
Thus any amount of mutual information 
(or connectivity) between two intervals reduces its entanglement entropy,
as we  note that $I$ appears with a negative sign
in \eqn{lko10}. Thus
with out any further thought it can be  said that the
entropy of connectivity between the bath pair is simply 
the mutual information
$
S_{connectivity}= I(B:B)$,
where mutual information
\bea
I(B:B)&=&{L\over 2G_3} \log ({ (2a+b)^2\over 2a l}) 
\eea
In 2-dimensions ${L\over 2G_3}={c\over 3}$, $c$ being the CFT  central charge.
We shall keep total size $l$ ($l=2a+2b$) fixed 
and would like to vary $b$, at the cost of $a$,
so as to study the evolution of entropy of the bath pair system 
as it grows in the size and find related 
mutual information quantities, see the numerical
plots in figures \eqn{fig24MI}
\& \eqn{fig24MIbh}. The bath entropy initially rises with  size $b$,
gains a maximum value and then falls to lower value. 
The mutual information of bath pair however
rises monotonically and follows entanglement entropy all along. 
Note that these plots
involve $S_{connectivity}$ contribution in  crucial way. Recently the
 connectivity contributions to entropy between  disjoint CFT patches 
appear in various ways in the literature, 
viz. islands,  Einstein-Rosen bridges, and the wormholes etc. 
Obviously in our case
here the connecting components get mainly assembled in  single  term
 $I(B:B)$ in eq.\eqn{lko10}.

The bath entropy in \eqn{lko9} has several competing components 
depending upon how large or small $b$ is compared to $a$. 
For example when $b\sim 0$, i.e. $2a\sim l$,
\bea
S[B]\simeq 2S(b)={L\over 2G_3} \log{b^2\over\epsilon^2}
\eea
precisely because  $I\approx 0$.
If $b$ increases,  $S[B]$  grows until the critical 
point is reached. At the criticality
\bea
{\partial S[B]\over \partial b}|_{b=b_c}=0\eea
The value $b=b_c$ is the maximum of bath entropy \eqn{lko9}. 
The critical point actually corresponds to well known Golden ratio
\bea 
{2a_c\over b_c}={b_c\over 2a_c+b_c}
\eea
which has a solution:
\bea
{2a_c\over b_c}={\sqrt{5}-1\over 2}\approx .62 \ .
\eea
It is similar to the natural Golden ratio of two successive 
numbers in the Fibonacci (Pingala) sequences 
$(0,1,1,2,3,5,\cdots ,34, 55, 89,...)$,
especially when bigger  numbers $34, 55, 89, ... $ 
in sequence are taken into consideration. 
The existence of entropy maxima implies that there exists
 a transition point for $S[B]$ where in
post transition period the entropy starts decreasing, see the plots in figure
\eqn{figent3a}. 
We study the evolving nature of system-B entropy
in these two regimes in greater detail  below.

\subsection{Small bath regime}
 When $b\ll a$,  in this regime of  small size bath systems we can write
\bea\label{phas1}
&& S[B]\simeq  
 {L\over 2G_3}\log{b^2\over \epsilon^2} 
- {L\over 2G_3}{b^2\over 4a^2}+{\rm subleading~terms},  
\eea
while mutual information of the bath pair is given by  
\bea
\br&& I(B:B)\simeq {L\over 2G_3}{b^2\over 4a^2}-{\rm subleading~terms},  
\eea
So $I$ remains negligible but nonzero for finite systems. 
One can easily  note from eq.\eqn{phas1} that for   
smaller $b$  the entropy is mainly determined by
the leading term as it gets  leading contribution from
two  RT surfaces ending on individual
bath intervals, each having width $b$ (symmetrical case).
The situation will be like this for any $b$  
so long as $b\ll a$. As $b$ becomes large while $a$ decreases, 
the bath entropy  grows, see fig.\eqn{figent3a}. Here it is vital to
note that the mutual information of bath pairs
never vanishes completely, no matter how small $b$ becomes or how  
large $a$ could have been!
As a consequence the mutual information  is never  zero!\footnote{
This discussion may remind us of  EPR paradox in classical relativity. 
The two separate spin systems in  vacuum 
would remain entangled sharing finite amount of (quantum) information 
between them no matter how far 
they might get separated. It is fair to say that $I\ne 0$ 
for  {\it finite} quantum systems.}
This fact is summarised through following inequalities
\bea
S[B]\le 2S(b), ~~~~~~~I(B:B)> 0
\eea
for finite disjoint bath pair systems. 
In conclusion, in $(1+1)$ dimensions 
the mutual information between two $CFT_2$ systems:\\
i) is positive definite and always a 
decaying function of the separation between two systems,\\
ii) is minimum when systems are farther separated,\\
iii) grows proportionally as the product of individual size
of the pair involved,\\
iv) and  is inversely proportional to the square of 
the separation between the systems $(d=2)$. 

Furthermore up to leading order
it can  be shown explicitly  that 
for an asymmetrically arranged bath systems, $B_1$ 
\& $B_2$, i.e. having different sizes,
the mutual information law becomes
\bea
I(B_1:B_2)={c\over 3} {b_1\times b_2\over D^2}+{\rm ~corrections}
\eea
where $b_1$ is the size of system-$B_1$ (on the left) and $b_2$ is the size of 
system-$B_2$ (on right). Here $D=2a$ is the actual separation between 
two  systems, and such that $b_1,~b_2\ll D$. 
 
As bath grows in size the subleading terms  become quite important.
Also  the net mutual 
information between them grows proportionately,
but it is inversely proportional to square of the distance between systems.
This phenomenon could be due to increased  ordering
between states (Hilbert spaces) of the bath systems as they come closer.
The increased ordering leads to overall reduction in the bath 
entanglement entropy. 
In other words  increased mutual
information leads to high correlations between the bath constituents. 
These strong
correlations (ordering) between systems can be only  at the cost of entropy. 
It is  well known entropic property of the physical systems. 

When $b$ approaches critical value $ b=b_c$  the
all subleading terms become very important. 
We should rather use exact formulae
as given above. Indeed there 
is a phase transition at $b=b_c$. The ever rising 
mutual information (due to  growth of 
correlations or connectivity between constituent bath pair) 
becomes the leading cause of this phase transition. At the transition point
the ratio $x={2a\over b}$ of system sizes satisfies the criticality equation 
\bea
x^2+x-1=0 .
\eea
The solution corresponds to
 critical ratio: $x_c={\sqrt{5}-1\over 2}$,
 which is also known as Golden ratio. 

\subsection{Beyond criticality: Islands}
 On the other side 
of  transition point, i.e. when $ b\gg 2a$, we 
find that the bath entropy starts decreasing and the
leading behavior is essentially given by
\bea\label{phas2}
S[B]
 &\simeq& S(l)+S(2a)- {2L\over G_3}
{a\over b}-\cdots
\eea
The $S(l)$ is entropy of full system $A\cup B$, 
it is also largest term,  and it is just  constant as
 $l$ is fixed. We vary only $a$ at the cost of $b$ keeping  $l$ fixed. 
The second term $S(2a)$ is of local nature. So when $a$ decreases,
it determines the fall of $S[B]$. The third term on rhs of eq.\eqn{phas2} 
is  subleading one and should be identified as 
  `island'   contribution \cite{hs2022, hs2024}. All the remaining
terms are also of local nature and these are crucial for understanding
of overall fall in the bath
entropy in the aftermath of transition point. 
At the same time the mutual information of the bath pair grows
steadily (which implies increased connectivity and 
strong ordering between the bath constituents), which is
\bea
I(B:B)&=&{L\over 2G_3}\left( \log{b\over 4a}+\log{(1+2s)^2\over 1+s}\right) 
\eea
So indeed in the large bath regime
 $I$ has a leading  expansion  
\bea\label{lko11}  
&& I(B:B)\simeq {L\over 2G_3}(\log {1\over 4s}+ 3s)-O(s^2)
\eea
where $s$ is  defined as
$s\equiv {a \over b}$. It  emphasizes that mutual
information of large symmetrical bath pair grows logarithmically 
as their physical separation $(2a)$ decreases. 
The  $3s$ term on rhs of eq.\eqn{lko11} is  subleading one and
  can be traced to island   contribution, precisely because
it falls off like  gravitational entropy of codim-2 
 surfaces (for island boundaries located at $z= b$) 
in the {\it near}-AdS ($NAdS_2$) gravity. 
Note that the gravitational entropy of island boundary \cite{almheri}
in $NAdS_2$ may be  expressed as 
\bea\label{lko11s}
S_{island}={\Phi_0\over 2G_2 b}. 
\eea
From \eqn{lko10}, \eqn{lko11} and \eqn{lko11s}, the bath entropy
\bea
S[B]=S(2b)+S(2a)-S_{island}+{\rm sub-subleading~terms}
\eea
where we should relate
\bea\label{aphi}
{2\pi R\Phi_0\over 3L}\equiv {a}, ~~~~~{2\pi R\over G_3}={1\over G_2}.
\eea
 Here $\Phi_0$ is  dilatonic boundary value of  near-AdS2  ($NAdS_2$)  
or 2-dimensional JT like gravity.
To understand this connection better one should  
resort to hybrid formulation of CFT together with gravity, i.e. `CFT2-NAdS2-CFT2' type  
construction of \cite{almheri}; see the details 
in the Appendix and also given in \cite{hs2024}. The  
sub-subleading terms, $O(s^2)$ and beyond, in eq.\eqn{lko11} 
are together called as `icebergs' terms! 
But these icebergs do not find simple geometric interpretation, other than possibly
by uplifting to two or more higher dimensional AdS spacetime.
(For example a $T^2$ compactification of an $AdS_4$ gravity, 
 would also produce $NAdS_2$  dilatonic backgrounds. But such terms are many!
The best strategy would be to resum them all, whenever possible.)
With  relations as in \eqn{aphi}, 
the icebergs may also be taken as higher order stringy
 corrections! At the perturbative level 
the island and icebergs are simply depictions  of variety of 
interactions between states of system-A and system-B (bath). 
In overall we learn that perhaps 
this is how mutual information and entropy of disjoint pair systems
may be constituted or partitioned
into various elements or components.

Thus the entropy of a larger bath can be expressed as
\bea\label{phas2nm}
S[B]&=& 
S(2b)+S(2a)
- {3L\over G_3}s +{7L \over 2 G_3}s^2 -{15L \over 3 G_3}s^3 
+\cdots
\eea
We have depicted this equation in the figure \eqn{figISLaw} for an illustration.
\begin{figure}[h]
\centerline{\epsfxsize=3in
\epsffile{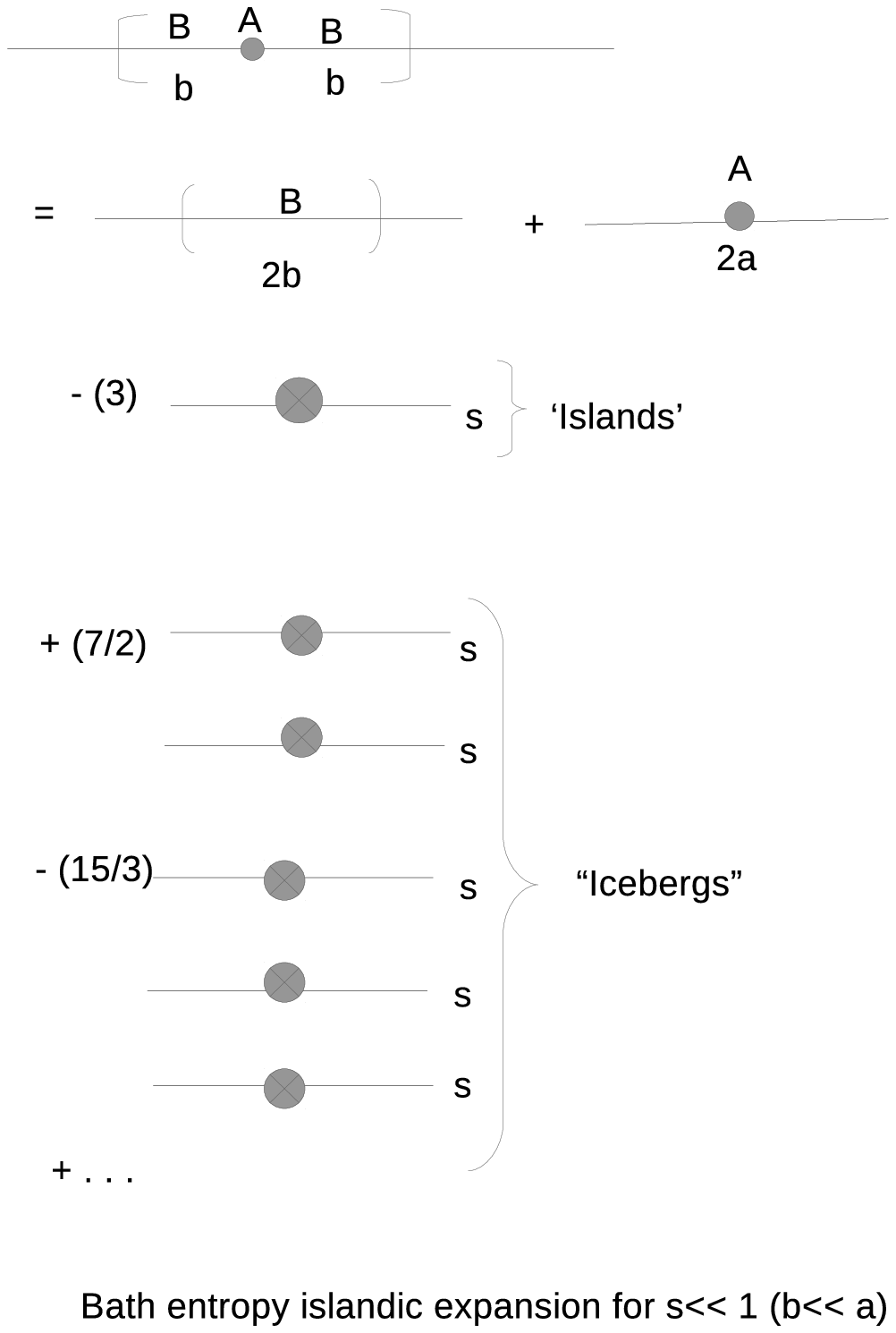}}
\caption{\label{figISLaw}
\small\it
 This sketch depicts the Taylor expansion of entropy of complete bath system-B 
when $s\ll 1$. Especially
the 3-rd term linear in $s$ is given  geometric interpretation 
in form of  gravitational island entropy. 
(E.g.  $s={a\over b}\ll 1$ and if we can take size $a\approx {\pi R}$).}
\end{figure}

Also we find that neither island nor the icebergs entropies remain
 significant towards the end of entropy
Page curve as these contributions usually 
decay much faster! Although  they do play significant role 
near the transition point (entropy maxima). In 
similar way the island and iceberg entropies are also
constituents of mutual information, but they
again remain subleading in post
transition regime, where the bath system has acquired a much larger size. 
From  leading terms in \eqn{lko11}
we observe that typically the relative entropy, $S(b)-S(2a)$,  
forms the leading contribution to mutual information, and hence to
the fall in bath entropy.

In the limit when $a\to 0+ O(\epsilon)$ (i.e.  $b\to {l\over 2}-O(\epsilon)$)
we have a safe limit for bath entropy
\bea\label{kkl3}
S[B]\to {L\over 2G_3} \log {l\over\epsilon}
\eea
as it should be. 
When the system-A in center nearly disappears,
the surrounding bath system-B will occupy the entire $A\cup B$ space (of net
size $l$). In that situation the entropy $S[B]$  ought to be 
given by \eqn{kkl3},
which is a measure of the entanglement of $A\cup B$ with the compliment system
$B^c$. 
In essence this establishes 
the Page curve of 2-dimensional CFT subsystem and surrounding bath 
at zero temperature.

\section {CFT systems in $d>2$} 
Let us discuss the entropy and Page transition in higher dimensional CFT also.
We  study only strip-like subsystems for simplicity. 
The entanglement entropy of strip shape system  can be found
 by obtaining areas of extremal RT surfaces \cite{RT,HRT}
\bea\label{fim1}
S(b)
&=& 
{L^{d-1} V^{(d-2)} \over (2d-4) G_{d+1}}\left( {1\over \epsilon^{d-2}}
-{2^{d-2}b_0^{d-1}\over b^{d-2}}\right)\br
\eea
where $b$ is the width of strip and $V^{(d-2)}$ is the 
regulated (box) volume of remaining $(d-2)$ spatial coordinates.
 
For small size  system-B
(being treated as bath on both sides) the leading extremal surfaces are
the disconnected ones. But there will be further corrections to it. 
The corrections arise because
there are contributions which come from other RT surfaces that
connect  end points of the opposite system of the bath pair 
in different ways.
 The net entropy of  system-B can be expressed as 
\bea\label{fin1}
S[B]&=& S(b)+S(b)-I(B:B)
\eea
where $S(b)$ is the entropy of the strip of width $b$ as given 
in \eqn{fim1}.
 The first two term in eq.\eqn{fin1} involves area contributions 
from two individual extremal surfaces anchored on respective $B$ system, 
refer to upper graph in figure \eqn{fig22b}.
The mutual information is obtained as
\bea\label{inf2}
I(B:B)&=& 2S(2a+b)-S(l)-S(2a)
\eea
which involves all other RT surfaces that connect the end points of opposite
pair in the system-$B$.\footnote{ This is similar to the CFT2 case where exact results are known \cite{cftENT}. 
We have assumed that $I(B:B)$ expression is correct for $CFT_d$ with $d>2$.
It would be true at least for strip systems cases, as these can be reduced to 2-dim by compactification.}  
There is a factor of 2 in first term due to two  identical 
shape RT surfaces are involved. We are using expression
 eq.\eqn{fim1} for all function $S(x)$. So
when $(b\ll a)$,
after some simplifications we may express,
\bea\label{inf03}
S[B]&\simeq& 2 S(b)-{\rm subleading~terms }\br
I(B:B)&\simeq& I_0{ b\cdot b\over D^d}-{\rm subleading~terms}
\eea
where distance parameter $D=2a$ and $I_0={L^{d-1}V^{(d-2)} 
\over (d-2) G_{d+1}}\equiv {c \over 3}$ is 
fixed proportionality constant, $c$ being the central charge.
The expression \eqn{inf03} indeed suggests 
a very simple law for mutual information between  
far separated pair of strips in  CFT$_d$. 
Thus for all higher $d$ cases, the mutual information between  pair of
strips falls off  rapidly as ${1\over (Distance)^d}$. 
While it is always proportional
to the product of respective system sizes going as $b\cdot b$.
Note $I(B:B)$ will never be zero for finite size systems, 
unless
 systems are placed at infinity; hence $I>0$, and $~~S[B]<2S(b)$. 
Thus at long distances mutual information
of far separated systems
 behaves much like asymptotically decaying law between two patches.

The entanglement entropy of  small size system-B will be defined by
  equation \eqn{fin1} approximately  far away from  
the crossover point, whereas $I$ remains negligible. 
Near  crossover point the $I(B:B)$ 
acquires  much more significance. 
There the area of RT surfaces connecting ends of opposite  system in the
bath pair become dominant; 
the leading RT surfaces  are drawn in lower graphs of figure \eqn{fig22b}.
In eq.\eqn{inf2}, 
the $I$ arises from the difference between  entropies
of lower two graphs. 
Note $I$ is a monotonic function and is devoid of 
UV divergences. 

\subsection{Critical points} 
We find that, for fixed $l$, while varying $b$
the critical point of bath entropy $S[B]$ in \eqn{fin1}
is reached when the relative size ratio $x$ $(x={2a\over b})$
 satisfies the following identity
\bea
(1+x)^{d-1}\cdot x^{d-1}= (1+x)^{d-1}-x^{d-1}
\eea
which may also be expressed as
\bea
{1\over x^{d-1}}-{1\over (1+x)^{d-1}}=1
\eea
That means the product of quantities $(1+x)^{d-1}$ and $x^{d-1}$ 
is the same as their mutual difference 
for some critical value  $x=x_c$. It has been checked that
the  critical point is always
an entropy maxima. Especially, as mentioned earlier,
for $d=2$ case the ratio $x_c$ comes as perfect Golden  ratio
as  in Fibonacci number sequence.

For $d=3$ CFT case the critical ratio follows an square-law
\bea\label{nlaw1}
 (1+x)^2\cdot x^2=(1+x)^2- x^2
\eea
We find that the corresponding critical value of ratio becomes $x_c\simeq .88$. 
See the figure \eqn{figsqlaw} drawn for an illustration. 

\begin{figure}[h]
\centerline{\epsfxsize=3.5in
\epsffile{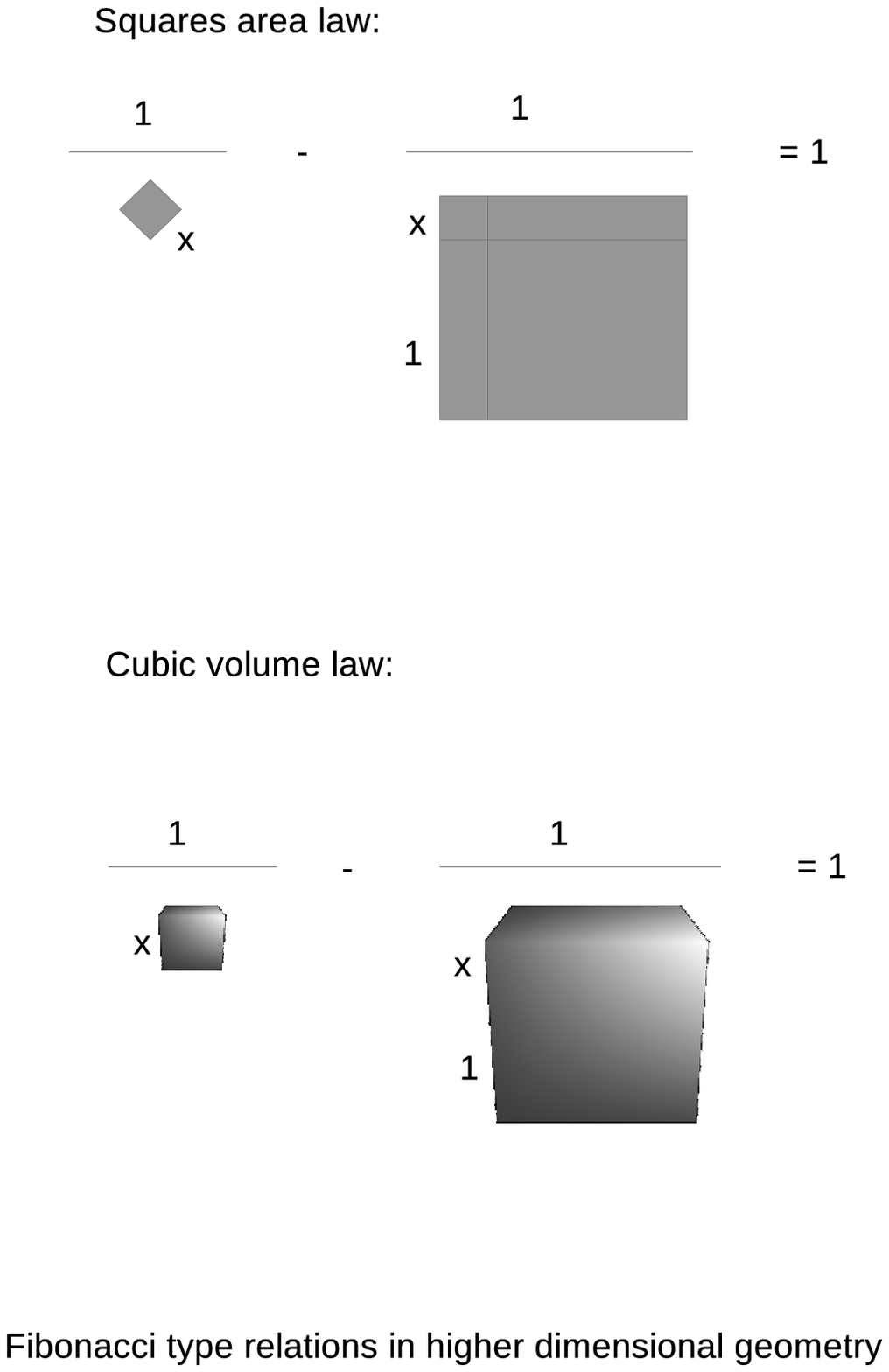}}
\caption{\label{figsqlaw}
\small\it
For $CFT_2$ the law governing proportions ${1\over x}- {1\over x+1}=1$ has 
a critical solution $x_c={\sqrt{5}-1\over 2}\approx .62$ which is the
Golden ratio as for Fibonacci or Pingala number sequences. Here it is shown
that there can be parallel analogues in higher dimensional geometry 
involving squares and cubes etc. }
\end{figure}

Similarly in $d=4$ the critical ratio follows a cubic volume-law
$$ (1+x)^3\cdot x^3=(1+x)^3- x^3.$$ 
for which the critical solution is $x_c\simeq .96$.

The above analysis suggests that in higher dimensional CFTs the critical ratio 
 $x_c\to 1$ for strip shaped subsystems. That means when bath size ($b$) 
on either side becomes as large as system-A $(2a)$ itself 
 the critical point arises and the bath system acquires maximum entanglement
entropy. 
Beyond the critical point $S[B]$ starts falling again. This 
is in essence  the Page curve realization!

\begin{figure}[h]
\centerline{\epsfxsize=2.5in
\epsffile{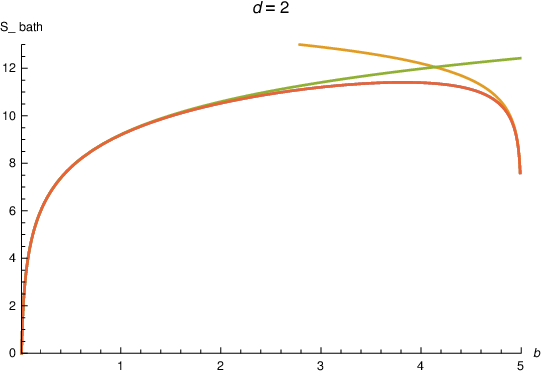} }
\caption{\label{figent3a} 
\small\it 
Entropy plots for the values $l=10,~b_0\ep=.01$, for  $CFT_2$ system. The  
upper falling curve in yellow (for $\sim S_l+S(2a)$) is  preferable for  entropy
in large $b$ region only $(b\gg b_c)$. The rising curve in green (for $\sim 
2S(b)$) is  good for entropy in the small size bath
region, i.e. $(b\ll 5)$. There is a crossover in between. 
The lower most graph (in red) is bath entropy for any $b$ and 
 is the physical one. 
It has a maximum at $b=b_c$. We have set $ {Lb_0\over 2 G_3}=1$ for the graphs.}
\end{figure}

\begin{figure}[h]
\centerline{\epsfxsize=2.5in
\epsffile{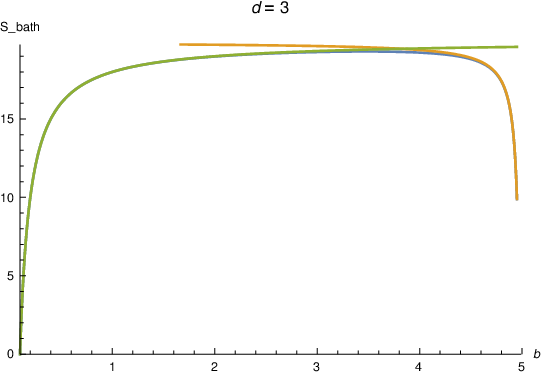} }
\caption{\label{figentcft3} 
\small\it 
Entropy plots for the values $l=10,~2b_0^2\ep=.01$, for  $CFT_3$ system. The  
upper falling curve (in yellow) is  preferable for  entropy
in large $b$ region only $(b\gg b_c)$. The rising curve (green) is  
good for entropy in the small bath
region $(b\ll 5)$ only. The  lowermost  continuous graph (blue) 
is bath entropy for any $b$ and 
 is actual one. 
It includes island contributions and has maximum at $b_c \sim 3.6$. We  set $ {L^2V^{(1)} b_0^2\over  G_4}=1$.}
\end{figure}

\begin{figure}[h]
\centerline{\epsfxsize=2.5in
\epsffile{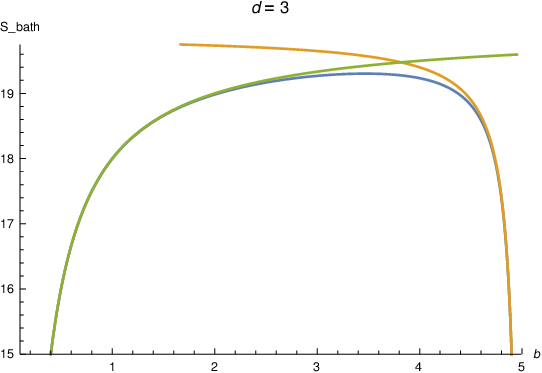} }
\caption{\label{figentcft3a} 
\small\it 
Entropy plots (zoomed out portion of figure \eqn{figentcft3}) for  $l=10,~2b_0^2\ep=.01$, for  $CFT_3$ system. The upper falling curve (in yellow) is  preferable for  entropy
in large $b$ region only $(b\gg b_c)$. The rising curve (green) is  
good for entropy in the small bath region $(b\ll b_c)$. 
The lowermost continuous graph (blue) is the bath entropy for any $b$ and 
 is the valid curve that includes exact island entropy. 
It has maxima at $b_c$. We set $ {L^2V^{(1)}b_0^2\over  G_4}=1$ for these graphs.} 
\end{figure}

\begin{figure}[h]
\centerline{\epsfxsize=2.5in
\epsffile{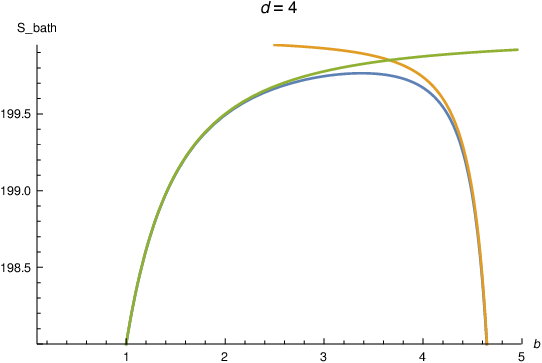} }
\caption{\label{figentcft4} 
\small\it 
Entropy plots for the values $l=10,~4b_0^3\ep^2=.0001$, for  $CFT_4$ system. The  
upper falling curve (in yellow) is  preferable for  entropy
in large $b$ region only $(b\gg b_c)$. The rising curve (green) is  
good for entropy in the small size bath
region only. The lowermost graph (blue) is bath entropy for any $b$ and 
 is the physical one and it includes island entropy.
There is  a maximum at $b_c\simeq 3.37$.
 We have set $ {L^3V^{(2)}b_0^3\over  G_5}=1$ for these graphs.}
\end{figure}

\begin{figure}[h]
\centerline{\epsfxsize=2in
\epsffile{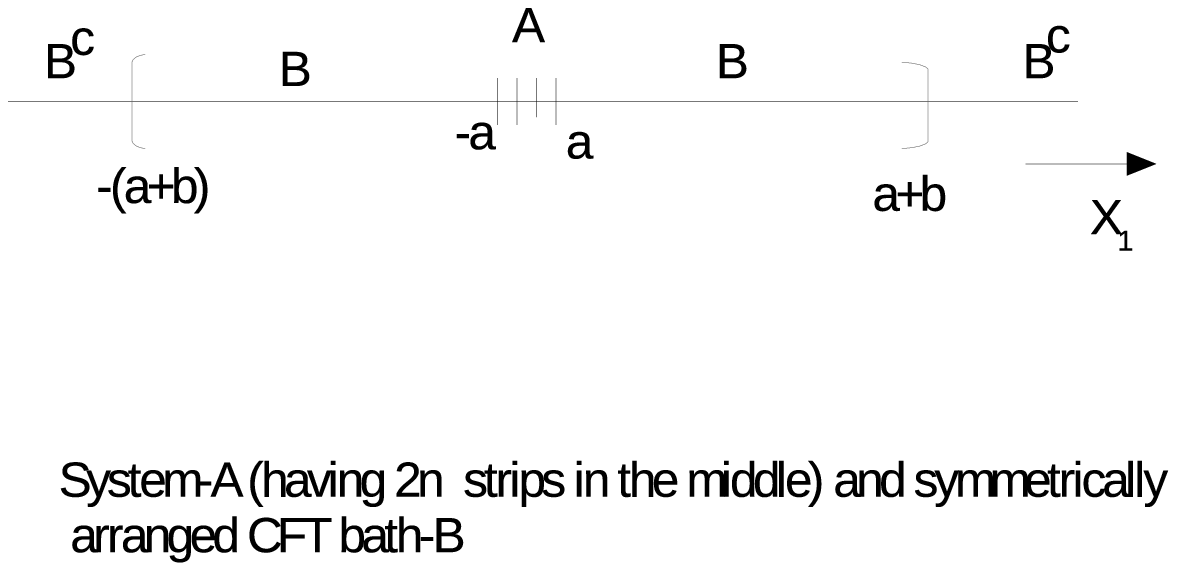} }
\caption{\label{fig21b} 
\small{\it An arrangement of system-A (in the middle on $x_1$-axis) and 
 CFT bath subsystem-B of size $b$ each lie on either side. All transverse spatial directions
(if any) except $x_1$ direction are suppressed. 
The complimentary system $(B^c)$ is also drawn.} }
\end{figure}

\section{Exact island scenario} 
So far we have discussed the island and icebergs entropy separately and approximately.
It is observed that beyond the critical point of the bath entropy, 
i.e. for sufficiently large bath pairs, 
it is always possible to rewrite the entropy \eqn{fin1} in following manner
\bea\label{islan1}
S[B] = S(l) + S(2a)- S_{island}(a,b)
\eea
The last island term  is exactly given by
\bea\label{gh67}
S_{island}=
{L^{d-1} V^{ (d-2)}\over
(d-2)G_{d+1}}
{2^{d-2} b_0^{ d-1}\over b^{d-2}}
(1-(1 + 2s)^{2-d})
\eea
which is valid for $d>2$ and is positive definite, 
i.e. $S_{island}\ge 0$. Unlike $S(l)$ and $S(2a)$, the contribution of
$S_{island}$ is free from UV divergences. Due to overall minus sign the
$S_{island}$ contribution to $S[B]$ is always negative, 
but importantly it can be exactly known!
The eq.\eqn{islan1} also implies that had we just taken $S[B]=S(l)+S(2a)$ then
there would have been an overcounting of bath entropy (entangled states). 
The inclusion of islands contribution reduces this excess counting of the entropy.
This fact may be summarised through an inequality
\bea\label{islan1ab}
S(l) -|S[B]- S[A]|\ge 0.
\eea
Once the island contributions are included 
the inequality \eqn{islan1ab}
 turns into exact equality given by \eqn{islan1}.
This finite systems inequality 
is quite analogous to  Araki-Lieb inequality 
for bi-partite thermal systems \cite{arakilieb}
\bea
S_{BH}-|S[A^c]-S[A]|\ge 0
\eea
where  quantum correlations 
(mainly due to bulk-to-bulk information) between far separated bulk systems 
 reduce the value of difference $|S[A^c]-S[A]|$, see  
\cite{faulkner}.
 
 Now using the relation:
 \bea
1-(1 + 2s)^{-n}=2s\sum_{\alpha=1}^n {1\over (1+2s)^\alpha} 
\eea
we can  express eq.\eqn{gh67} as
\bea\label{gh68}
S_{island}=
{L^{d-1} V^{ (d-2)}\over
(d-2)G_{d+1}}
{2^{d-1} b_0^{ d-1}a\over \bar b^{d-1}}
\eea
where $\bar{b}$ is an exact location of island boundary and is given by
\bea
{1\over \bar b^{d-1}}={1\over b^{d-1}}
\sum_{\alpha=1}^{d-2} {1\over(1 + 2s)^\alpha}.
\eea
The right hand side of the eq. \eqn{gh68} now 
behaves like gravitational entropy.
For better understanding it will be useful to  analyze the 
island entropy on case to case basis. 

{\bf $d = 3$ CFT case:}
Particularly in 3-dimensional case the island entropy takes the form
 \bea
S_{island}=
{L^2 V^{ (1)}\over G_{4}}
{2^2 b_0^{2}a\over b^{2}}{1\over (1 + 2s)}\equiv
{L V^{ (1)}\over 2G_{3}}
{\Phi_0\over \bar b^2}
\eea
where we have related $a \equiv {\pi R\Phi_0 \over 2^2 Lb_0^2}$, 
and 3-dimensional Newton's constant is given by $G_3 = G_4 /(2\pi R)$.
These relations would automatically arise if we consider 
 $ S^1$ compactification of $AdS_4$ geometry. 
(The $R$ is radius of 
compactification circle, and $\Phi_0$ is the dilatonic constant 
in lower dimensional  $NAdS_3$  gravity.)  
Here $\bar b$ provides exact location of the 
island boundaries (two boundaries in symmetric case) in  $NAdS_3$ geometry, that is
\bea\label{kin2}
z_{island}=\bar b,
~~~~~\bar b= b \sqrt{1 + 2s}
\eea
here $z$ being holographic coordinate of $NAdS_3$ spacetime. 
Since the ratio $s \equiv {a\over b}\ll 1$,
one indeed gets
$\bar b \simeq  b + O(s)$
 approximately. 
Hence the exact location of  islands boundary is not far away from $z=b$,
in   $s\ll 1$ post crossover regime. 
Also the mutual information of bath system has 
a simple series expansion too,
\bea
&&I(B:B)={L l_2\over 2G_4 a}\sum_{n=0}^\infty (-1)^n C_n s^n
\eea
where $C_0=1, ~C_{-n}=0$ and other coefficients are known through
the relation 
\bea
&& C_n=2C_{n-1}+1, ~~~~
\eea
Thus mutual information will diverge whenever separation ($2a$) becomes  small enough. 
This is consistent with general expectations,  that the
mutual information would be maximum when two systems become closer to each other
 and it is minimum when they are  situated far apart and being smaller.

{\bf  $d = 4$ CFT:}
The island entropy  in $CFT_4$ case becomes
 \bea
S_{island}=
{L^3 V^{ (2)}\over 2G_{5}}
{2^3 b_0^{3}a\over b^{3}}{1+s\over (1 + 2s)^2}\equiv
{L^2 V^{ (2)}\over 2G_{4}}
{\Phi_0\over \bar b^3}
\eea
 where we related  $a = {\pi R\Phi_0 \over 2^2 Lb_0^3}$, 
and 4-dimensional Newton’s constant being $G_4 = G_5 /(2\pi R)$.
These would arise once we consider  $S^1$ compactification of $AdS_5$ 
along $x_1$ (say). Here  
$R$ is the radius of compactification, 
and $\Phi_0$ is suitable dilatonic constant in $NAdS_4$ geometry.
The $\bar b$ gives an exact location of the 
 island  boundary inside
the $NAdS_4$ bulk,
\bea
z_{island}=\bar b,~~~~~~\bar b= b {
(1 + 2s)^{2\over 3}\over (1 + s)^{1\over 3}}\sim b (1+s+O(s^2))
\eea
It will happen when we take  hybrid $CFT_4-NAdS_4-CFT_4$ set up, 
see in Appendix.
Therefore we can conclude from above two cases that an exact 
island location can be determined for all $d>2$ cases upon resumming the
subleading icebergs terms. 
This has been possible at least for strip shape CFT
systems. For the closed subsystems, e.g. circle or sphere,
 the surrounding bath system is always remains a connected system.

We have  plotted bath entropy for CFTs in various dimensions in the figures 
\eqn{figent3a}
\eqn{figentcft3}
\eqn{figentcft3a}
\eqn{figentcft4}. These are for illustration purpose only. All the graphs have
a critical point with an entropy
maxima. In small bath region $(b\ll l)$ the bath entropy is
approximately given by $\sim 2S(b)$. While in the large bath regions the 
entropy is approximately  governed by
 $\sim S_l+S(2a)$. However in the near critical
 region there are significant island corrections, so full expansion
 $S[B]=S_l+S(2a)-S_{island}$ becomes important. The continuous curves
are exact plots of bath entropy.

\subsection{Tri-partite information}

The tri-partite information $I_3(A:B_1:B_2)$ is yet another measure of information
of system-A when entangled with system-$B_1$ and  system-$B_2$ individually as
compared to case when $B_1$ and $B_2$ are considered together ($B_1\cup B_2$). 
It is defined as
\bea
I_3(A:B_1:B_2)= I(A:B_1) +I(A:B_2) -I(A:B_1\cup B_2) \le 0
\eea
Essentially it is a measure of extensivity 
of the information of system $A$. Especially
for the cases  when all systems are adjacent to each other $I_3=0$.
So that there is  complete fidelity, and no missing information,
 when all systems are adjacent.
For our set up of central system-A and the adjacent symmetrical
 pair making the system-B
 we  find that $I_3(A:B:B)=0$.

\subsection{ Local entropy functional}
   
It is evident that in the post transition regime
 described by eq.\eqn{islan1}, 
we have a well defined local entropy functional $S_{local}(a,b)$ such that
for suitably large $b$ the 
entropy can be expressed as $S[B]= S_l +S_{local}(a,b)$, where local part is
explicitly 
\bea
S_{local}(a,b)\equiv S(2a)-S_{island}(a, b) .
\eea
The  $S_{local}$ is having all local information about systems A and B, 
whenever there is any small change in their sizes. While
$S(l)$ is  constant because the entire system set up is conserved, 
and $l$ has a fixed value. 
So in the large bath regime, $S_l$ does not play  
significant role other than being a
fixed global component of the bath entropy. 
\footnote{ This is in consonance with
 quantum entropy of bath defined  using 
 minimality principle \cite{hs2024}. Thus if we wish to define quantum bath entropy
here, then $S[B]^{quantum}\equiv S_{local}(a,b)$. Alhough
 we do not do that here, but the island entropy
is inevitable even if we drop $S_l$.
} There is an obvious  bound $0\le S_{local}(a,b)\le S[A]$.
In other words we can explicitly break the large bath entropy into two parts
\bea\lll{minp3}
S[B]=
S_{global}[B]+S_{local}[B]\eea
where
$
S_{global}\equiv S_{l}.$ 
It is the local entropy part which is susceptible to any miniscule 
changes $(\delta a$) in the systems at the interface, 
when system-A is much smaller than the surrounding system-B. 
Note this conclusion is independent of how large $l$ could 
have been to begin with.  The  local (quantum) functional
$S_{local}(a,b)$ is complete  and it has all necessary local 
information about the entanglement between systems A and B,
 including information about their criticality! However the  
 local entropy  is defined such that it is
 devoid of constants in the definition.

Under $l \to \infty$ limit, keeping $a$ fixed, we find that
$S[B]\to S_l$ and the entropy become infinitely large, 
whereas  local entropy functional
$$S_{local}(a,b)\to  S[A]$$
thus remaining finite! Hence the
$S[B]_{local}$ 
becomes equal to the entropy of system-A in  strict $l\to\infty$ limit.
The net bath entropy goes as $S_\infty+S[A]$ as expected. 
(For system in pure state $S_\infty=0$).

\section{KK interpretation and island corrections}

It is enough to discuss  special case of 
CFT3 systems described by  eq.\eqn{islan1}. 
 Consider   small size for system-A, such that $a\simeq 2\pi R$, 
where $R$ is  Kaluza-Klein scale of the theory 
(analogous to JT-gravity and near $CFT_1$ theory).
(We expect that for such small size system-A with narrow width, the
system-A 
could be effectively treated point-like, and as being compactified on $S^1$, 
with a compactification
radius $R$, while $R\gg\epsilon$.) In that case
we can safely take $a\simeq{\Phi_0 \pi n R\over 2b_0^2L}$, with $n$ being 
 positive integer. 
\footnote{Note, we have assumed that there is 
intermediate Kaluza-Klein compactification
on $S^1$ when  the 
size $a$ becomes approximately  $\simeq  R$.
It becomes plausible to study a  
lower dimensional dual gravitational description 
for system-A (like multiple strips of narrow width $ \pi R$ 
wrapped along circle  direction but
 fully extended along other transverse directions).}
For simplicity we only consider small 
 $n$ values,  also $\epsilon\ll n \pi R\ll b$
(If there is any difficulty one can simply take $n=1$). The 
system-A  can essentially be treated as  assembly of $2n$  
narrow (parallel) strips
sandwiched between symmetrical  system-B on either sides (
of individual size $b$). We have a situation
as depicted in the figure \eqn{fig21b}, 
where transverse directions $x_2$ and $x_3$ are all suppressed.  
An expression
on the r.h.s of \eqn{islan1}, for  $ { n \ll {b\over \pi R}}$, 
gives bath entanglement entropy exactly (for $d=3$)
\bea\label{fin2an}
S[B]
&&\equiv 
S_l +S_{2n-strips}[A]-S_{island}\br 
&&=S_l  +{L^2 l_2 \over 2 G_4}({1\over\epsilon}-{2b_0^4L\over n \pi R\Phi_0}) 
-S_{island}(\bar{b})
\eea
where leading term $S_l$ is fixed entropy of the
entire system and 
\bea
&&S_{island}
= {L l_2\over 2 G_3} {n\Phi_0\over \bar b^2}  
  \eea
  where $\bar{b}=b\sqrt{1+2s}$, as in \eqn{kin2}.
The $G_3=G_4/(2 \pi R)$ is 3-dimensional Newton's constant.
The islandic contribution, especially for $n=1$, 
is similar to the gravitational entropy
of an island boundary situated at $z=\bar b$ inside $NAdS_3$ bulk. 
Its contribution falls as ${1\over \bar b^2}$ as the size of bath enlarges.
As we explained $z=\bar b$ is the 
exact location of codim-2 boundaries inside the $NAdS_3$ bulk. 
There are no iceberg entropies here because $S_{island}$ 
is a resummed quantity and it includes all subleading contributions. 
If $\bar b$ in eq.\eqn{fin2an} is expanded, it would generate 
an infinite series, which of course is perturbative. 
Actually in these examples it will not be wise to separate icebergs from 
 other leading terms in \eqn{fin2an}, as
the exact location of the island surface is obtainable
upon resummation.
 It is  clear from the starting line of eq. \eqn{fin2an} itself. 
The hierarchy of various contributions on r.h.s. is as follows
\bea\label{fin2ank} 
S_l \gg S[A]_{2n-strips}\gg S_{island}
 \eea
where $b\gg\pi R$.
While  entanglement entropy of  $2n$ parallel
strips, all assembled side by side and in contact with bath, 
 is simply (for $d=3$)
\bea\label{fin2an2}
 S[A]_{2n-strips}=
{L^2 l_2 \over 2 G_4}({1\over\epsilon}-{2b_0^4L\over  n \pi R\Phi_0})
\eea

\subsection{Entropy spectrum of multi-strip systems}
We are  interested in exchanging  small number of
 strips (of CFT matter) between bath (B) and the multi-strip system (A).
This will lead to changes in the systems entanglement entropy. 
Note a small number of strip exchanges between B and A will 
not change the total system entropy $S_l$.
Thus net change in entropy of system-A
with KK-strips number $n_2$ and with KK-strips number $n_1$ 
can be found to be  $(n_1<n_2)$
\bea
\bigtriangleup S_{1\to 2}&&=S_{strips}[n_2]-S_{strips}[n_1]\br
&&\simeq
{L^2 l_2 \over  G_4}({1\over n_1}-{1\over n_2}) {b_0^4L\over \pi R\Phi_0}
+ {L^2 l_2\over  G_4} {2(n_1-n_2)\pi R\Phi_0\over L b_0^2  b^2} 
\eea
where second term has island contributions. 
 The  spectrum  will obviously discretized  due to KK scale!
A fixed quantum of energy (matter) exchanges is required to take place
between bath-B and  strips system-A during exchange process 
(of thin strips). It is one of the quantum
processes.
Note we are discussing CFT ground state at zero temperature only.
 This exchange process entails KK-level `jumps'. 
Here $n_1=1$ can be treated as the lowest level 
(smallest size single strip system)
while  $n_2>1$ will correspond to
 higher levels (more than one strip cases).
Note a change in the strip number
is necessarily associated with discrete 
(quantum) CFT matter exchanges between system and the surrounding bath. The 
relevant physical scale is set by compactification radius $R$.

\section{Summary and discussion}
We proposed that  entropy of entanglement for 
large symmetrical system-B  in contact with  smaller central system-A
can be written as
$$ 
S[B]= S_{global}(l)+S_{local}(a,b)
$$
where $S_{global}=S(A\cup B)$ is  full system entropy that has a
fixed value and it makes the 
global part. The functional $S_{local}(a,b)$ is local part 
and has all information about the systems
including their criticality and cross over (transition) points.
Any small fluctuation in size of system-A ($\delta a$)
due to quantum matter exchanges with system-B (bath)
will not alter this conclusion, as the
entire systems set up is conserved. 
The equation is also a realization of the entropy Page curve 
for  CFT matter in contact with a bath system. The bath entropy falls immediately
after the transition point where the island effects become significant. 
These effectively bring down the bath entropy. 
For all purposes $S_l$ is 
conserved and the information it contains is essentially global in nature.  
Furthermore the local entropy functional is given by
$$S_{local}(a,b)\equiv S(2a)-S_{island}(a,b)$$
where $S_{island}$ can be exactly known for $d>2$.  
We have explicitly shown  how islands and subleading iceberg
entropies contribute to $S_{local}(a,b)$ in post cross over regime. 

Especially it is also shown that for strip shaped systems  
exact islands can be defined by resumming the subleading icebergs entropies. 
It is true for  $CFT_d$ strip systems with $d>2$. At criticality 
where the bath entropy has a maxima, the ratio of systems sizes
follows relations which are Fibonacci type generalizations
 involving higher dimensional geometries
 $${1\over (x)^n}-{1\over (1+x)^n}=1.$$
like squares and cubes. Note $n=1$ case corresponds to the Fibonacci spiral.
  
  Since exact island
contributions can be known we could write an identity involving
equality of differences
\bea
S[B]-S[A]=S_l-S_{island}
\eea
whenever the bath system-B is sufficiently larger than system-A. 
Since this is true at every point after the cross over, 
if the bath size increases any further the changes 
in  $(S[B]-S[A])$
would have to be compensated precisely by the islands entropy only. 
In the $l\to\infty$ limit, keeping $a$ fixed,
one gets the desirable relation;  $S[A^c]-S[A]=S_{\infty}$. Note
 $S_{\infty}=0$ for pure state. For finite temperature systems
 the difference will lead to thermal entropy $S[A^c]-S[A]\simeq S_{BH}$.
 
Further it is found that mutual information for far separated  $CFT_d$ subsystems
follows an asymptotic law:
$$I(B_1:B_2)\propto {b_1.b_2\over D^d}$$ 
although the right hand side has subleading corrections to it. This however shows
that mutual information is never vanishing for finite systems
even if they are far separated, i.e. $D\gg b_1, b_2$.  

Furthermore we have  analyzed our results 
when central system-A  becomes point-like, 
i.e. becomes similar in size to the Kaluza-Klein
scale of theory, if there exists a compactification scale. 
When the system-A size approaches KK-scale, 
we expect necessary discreteness in the entropy. Should 
there be no $S^1$ compactification scale, the quantum entropy of  
large bath will be equal to $S_l$ when
  system-A disappears totally, or whenever $S_{local}\to 0$. 
 In summary it is observed that the
changes in bath entropy would capture Kaluza-Klein discreteness. 
 
 
\vskip.5cm
   
\appendix{

\section{Effective construction of  hybrid gravity and CFT systems}

For  small size central CFT system-A, such that its
size $a\approx 2\pi R$, i.e. when the system size can approximately 
fit within the Kaluza-Klein
radius, the system-A may be treated
as being point-like. The symmetrical bath subsystem-B on either side is being comparatively  
very large, so it can continue to 
be described by respective (noncompact) $CFT_d$. 
However, for all practical purposes, with out any loss of
physical picture, the 
system-A can also be replaced by dual `near-AdS' geometry living in one step lower
$d$ spacetime dimensions. 
The Newton's constant for near-AdS  geometry
would be $G_{d}=G_{d+1}/(2\pi R)$. We have tried to 
draw these situations in the figure \eqn{figeent23Pen} for systems $A$,
$B$ and the compliment CFT system $B^c$. (Note $B$ and $B^c$ together make an infinite bath system.)
The mixed gravity and CFT set up 
has been a favorable arrangement for the first
island proposal \cite{almheri}.

 \begin{figure}[h]
\centerline{\epsfxsize=3.2in
\epsffile{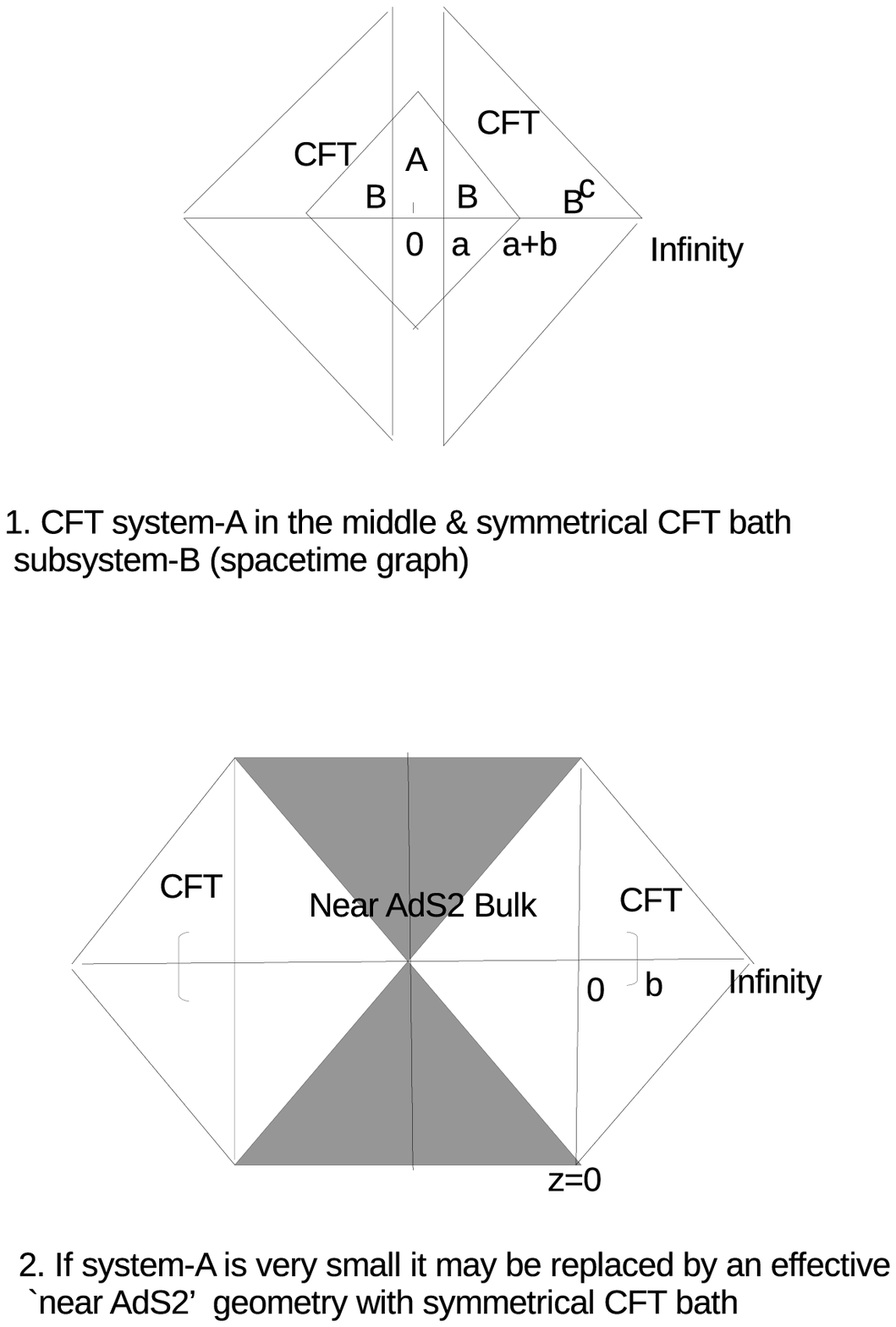} }
\caption{\label{figeent23Pen} 
\small\it An hybrid arrangement of  `near' $AdS_2$ gravity and $CFT_2$ systems. 
The gravitational constant $G_2$ may be suitably fixed. The AdS space divides CFT in two halves.
 The set up would be similar for all $CFT_d$ strip systems cases.  }
\end{figure}

The entropy of  gravitational island boundary located, say at $z=b$ (symmetrically),
is given by
\bea
\label{j7t9}
S_{island}={L l_2 \over 2G_3}{\Phi_0 \over b^2}
\eea
taking an example of d=3 case. The $\Phi_0$ is suitable
scalar parameter and is related to
the boundary value of the running dilaton field, 
$e^{-2\phi}={\Phi_0\over z}$,
 of the near-$AdS_3$ spacetime: $$ds^2_{3}
={L^2\over z^2}(-dt^2+dx_2^2+dz^2)$$ 
The islands are supposed to be located at $z=b$, 
and situated symmetrically on either side of the center 
in the gravity region in the figure \eqn{figeent23Pen}. 
Further, without difficulty 
we can take $\Phi_0={L b_0^2 a\over 4\pi R}$. Note $2a$ is the width of the CFT system-A
in the middle.
Thus hybrid gauge-gravity configurations can provide a
physical understanding of the island's contribution in the bath entropy
\eqn{islan1} in post Page transition regime.

A generalisation for $NAdS_d$ geometry is
$$ds^2_{d}
={L^2\over z^2}(-dt^2+dx_2^2+\cdots+dz^2)$$, with
 a running dilaton $\phi$, such that the scalar
$$\Phi\equiv e^{-2\phi}={\Phi_0\over z},$$
and the island entropy becomes
\bea
\label{j7tt9}
S_{island}={L^{d-2}  V_{(d-2)} \over 2G_d}{\Phi_0 \over b^{d-1}}
\eea
We may generically set $\Phi_0\sim {L b_0^{d-1} a\over 2\pi R}$ which has same mass dimensions
as $z$. (Note one  may 
have to tune final relations with  factors of 2 and $\pi$.)

}

%

\end{document}